\documentclass[aps,prd,amssymb,superscriptaddress]{revtex4}
\usepackage{graphicx,bm,color,psfrag,hyperref}
\usepackage{amsmath}
\usepackage{amssymb}
\usepackage{amsfonts}
\def\ar{ \!\!\!}

\begin{document}

\title{Accelerated expansion of the universe \`{a} la the Stueckelberg mechanism }

\author{\"{O}zg\"{u}r Akarsu}
\email{oakarsu@ku.edu.tr}
\affiliation{Department of Physics, Ko\c{c} University, 34450 Sar{\i}yer, {\.I}stanbul, Turkey}

\author{Metin Ar{\i}k}
\email{metin.arik@boun.edu.tr}
\affiliation{Department of Physics, Bo\u{g}azi\c{c}i University, 34342 Bebek, {\.I}stanbul, Turkey}

\author{Nihan Kat{\i}rc{\i}}
\email{nihan.katirci@boun.edu.tr}
\affiliation{Department of Physics, Bo\u{g}azi\c{c}i University, 34342 Bebek, {\.I}stanbul, Turkey}

\author{Mehmet Kavuk}
\email{mehmet.kavuk@boun.edu.tr}
\affiliation{Department of Physics, Bo\u{g}azi\c{c}i University, 34342 Bebek, {\.I}stanbul, Turkey}

\date{\today}

\begin{abstract}
We investigate a cosmological model in which the Stueckelberg fields are non-minimally coupled to the scalar curvature in a gauge invariant manner. We present not only a solution that can be considered in the context of the late time acceleration of the universe but also a solution compatible with the inflationary cosmology. Distinct behaviors of the scalar and vector fields together with the real valued mass gained by the Stueckelberg mechanism lead the universe to go through the two different accelerated expansion phases with a decelerated expansion phase between them. On the other hand, in the solutions we present, if the mass is null then the universe is either static or exhibits a simple power law expansion due to the vector field potential.
\end{abstract}

\maketitle

\section{Introduction}
\label{Intro}

The accelerated expansion of the universe came under scrutiny right from the inception of the concept of inflationary cosmology \cite{Starobinsky80,Guth80,Linde82,Albrect82}. Inflation is characterized by an epoch of accelerated expansion in the very early universe ($\sim 10^{-35}$ seconds) at energy scales $\sim 10^{16}\,{\rm GeV}$. It is not only the most prominent attempt to resolve the problems of standard Big Bang cosmology such as horizon and flatness problems, but also provides an elegant mechanism for the origin of large scale fluctuations in the cosmic microwave background (CMB) (see \cite{Linde14} for a recent review). However, a concrete and unique realization of inflation from a fundamental theory such as string theory \cite{Quevedo02} is still an illusive task. It has many variations usually based on general relativity (GR), where the inflation is driven by a scalar field(s) that is/are usually introduced in an ad hoc way. One may see \cite{Encyclopaedia} for a comprehensive list of scalar fields considered in the context of inflation.

Besides the early acceleration, it is today well established with independent studies \cite{SCP10,SDSS10,WMAP09,PlanckXVI} that the current universe is evolving with an accelerated expansion that started approximately 6 Gyr ago. We lack a satisfactory explanation for this current acceleration that happens at energy scales $\sim 10^{-4}\,{\rm eV}$, where we supposedly know physics very well. The most successful cosmological model accommodating this fact, we know so far, is the $\Lambda$CDM model based on GR. However, it suffers from two important theoretical problems known as the fine tuning and coincidence problems related with the cosmological constant $\Lambda$, which is mathematically equivalent to the conventional vacuum energy and is responsible for the acceleration of universe in this model \cite{Zeldovich,Weinberg89,Sahni00,Peebles03}. The latest data from the Planck CMB experiment, whose major goal is to test this model to high precision and identify areas of tension, shows a remarkable consistency with the predictions of the base $\Lambda$CDM model. However, it reveals also a number of intriguing features of the data that might be ascribed to the cosmological constant assumption of the model; for instance, it is found that the data alone is compatible with $\Lambda$ assumption, but a dark energy component yielding a time varying equation of state (EoS) parameter is favored when the astrophysical data is also taken into account \cite{PlanckXVI}. This is in line with the idea of describing dark energy as a scalar field that was first considered to alleviate the theoretical problems related with $\Lambda$. However, the scalar field models of dark energy are also mostly ad hoc and/or considered phenomenologically rather than being derived from a fundamental theory (see \cite{Copeland06,Bamba12} for comprehensive reviews on dark energy).

Scalar fields are in fact ubiquitous in theories beyond the standard model such as string theory and super-symmetry. The discovery of Higgs boson \cite{CMS,Atlas} with a mass $125\;{\rm GeV}$ has largely affirmed the existence of the Higgs scalar field and arose the interest in the possible existence of scalar fields with a mass consistent with the cosmological scalar fields, namely, the inflaton and dark energy fields. An alternative way of inducing accelerated expansion rather than using inflaton or dark energy sources in GR is to consider modified theories of gravity \cite{Nojiri10,Capozziello11extended,Clifton12}. Scalar-tensor theories are the most established and well studied modified theories, and appear at low-energy limits of string theories. Brans-Dicke theory of gravity \cite{Brans-Dicke} is the prototype of these theories. It involves a scalar field as an extra field mediating the gravitational interaction, while a scalar field is introduced as an external energy source (namely, energy-momentum tensor) in GR, which is a pure tensor theory of gravity. Moreover, the scalar field in Brans-Dicke theory couples directly to the scalar curvature, and gives rise to a dynamical effective gravitational coupling. Therefore, one should face with the possibility of a time dependent gravitational coupling in cosmological models based on such modified theories.

The only long range interaction which could be relevant on cosmological scales apart from gravity is the electromagnetic field, which is a vector field. A vector field based inflationary cosmological model was also suggested in 1989 \cite{Ford89} but it started to receive keen attention only a decade ago. In recent years, on the other hand, vector fields have been discussed and considered with an increasing interest not only as an alternative to the scalar field models of inflaton but also that of dark energy \cite{Koivisto06,dimopoulos07,Bamba08,Golovnev08,Koivisto08,Kanno08,Jimenez09a,Watanabe09,Jimenez09b,Kanno10,golovnev10,Thorsrud12,Bartolo13b}. The primary reason behind this increased interest is the efforts to explain some of the anomalies found in the large-scale CMB temperature in the WMAP data \cite{WMAP7an}. These anomalies have also been confirmed by the recent high precision Planck data \cite{PlanckXVI,PlanckXXIII,PlanckXXVI}. However, vector field models that give an accelerated expansion usually suffer from ghost instabilities \cite{Himmetoglu09a,Himmetoglu09b,esposito10} due to imaginary (tachyonic) mass. In particular, such inflationary models require huge mass for the vector field, and hence a huge amount of tachyonic mass which makes the issue even worse. In a recent study \cite{Akarsu14}, a cosmological model where Brans-Dicke gravity coupled to a vector field with a variable mass, which was constrained to be positive and real so that tachyonic mass was avoided from the beginning, was given. However, it was shown that in this case the vector field behaves like a dark matter source, and it is the scalar field which drives the accelerated expansion.

Motivated by the above discussion, in this paper we investigate a cosmological model where the Stueckelberg fields couple directly to the scalar curvature in a particular way. The reason being that Stueckelberg action \cite{stueckelberg38a,stueckelberg38b} involves both scalar and vector fields, and also such actions arise naturally in compactifications of higher-dimensional string theory \cite{ruegg04,Kors05}. Vector field actions with a mass term usually spoil the gauge invariance as in the Proca action that gives Maxwell's equations when the mass is set to zero. Stueckelberg \cite{stueckelberg38a,stueckelberg38b}, on the other hand, described a massive photon by maintaining gauge invariance by introducing a scalar field $B$ that mixes with the electromagnetic field $A_{\mu}$ under gauge transformations. The scalar field arises from the extra degrees of freedom and corresponds to the longitudinal mode of the photon polarization \cite{ruegg04,Kors05}. Extending this idea (i.e. stueckelberging the electromagnetic $U(1)$ and thus giving a mass to the physical photon) to cosmological scales and investigating cosmological solutions by constraining the mass term to positive real values is quite appealing. For instance, in a recent study \cite{Jimenez13} it is showed that the Stueckelberg fields can play the role of dark energy since they can give an effective cosmological constant on large scales.

We are particularly interested in the background expansion history of the universe and hence for convenience we consider spatially maximally symmetric and flat Robertson-Walker space-time. In accordance with this, assuming the universe is electrically neutral we consider only the temporal electromagnetic field i.e. the electric potential of the vector field. Temporal electromagnetic field \cite{jimenez10} and vector fields \cite{picon04,bohmer07,Koivisto08} are considered in the cosmological context. We follow the same approach to construct the gravitational action and treat the scalar field as the Jordan-Brans-Dicke (JBD) scalar \cite{faraoni04,arik05}. We propose new type of gauge invariant coupling to the scalar curvature applying a particle physics approach and investigate its cosmological solutions. The force mediated by a massive particle is given by the Yukawa type behavior $\sim \frac{e^{-mr}}{r}$, where $m$ denotes the mass. The laboratory bound on the photon mass is $10^{-14}\;{\rm eV}$, derived from the measurements of deviations from the
Coulomb law (i.e. $m=0$) \cite{coulomb71} potential, and is far above the bounds obtained from the astronomical and cosmological tests. The bound on $m$ is $\sim 10^{-15}\;{\rm eV}$ from the measurements of Earth's
magnetic field \cite{earth} and Pioneer-10 measurements of
Jupiter's magnetic field \cite{pioneer}, and is $10^{-27}\;{\rm eV}$ from the galactic magnetic fields
\cite{lakes98,vectorgalactic} (see \cite{goldhaber71} for a review). One may note that the higher the scale the tighter the bounds, which demonstrates also that even an extremely small value of the photon mass can have a considerable effect on the evolution of the universe. There are various applications of the Stueckelberg mechanism in the context of cosmology; for instance, it has been used as a natural source to account for some sort of dark matter related to the gauge-group parameter in \cite{aldaya06} and as a mechanism for giving a mass to graviton in the context of massive gravity in \cite{heisenberg11, heisenberg14}. In this study, on the other hand, we show that a massive non-minimally coupled photon that gains its mass by the Stueckelberg mechanism in curved spacetime may give rise to interesting expansion histories for the universe, even a history that can be considered in the context of inflation (including a switch-off mechanism) in the early universe and to the current acceleration of the universe.

\section{The gravitational field equations}
The action we propose is
\begin{eqnarray}
S&=&\int {\rm d}^{4}x\sqrt{-g}\bigg[-\frac{1}{8\omega m^2}\left(m B+\nabla_{\mu} A^{\mu}\right)^2R-\frac{1}{4}F^{\mu\nu}F_{\mu\nu}+\frac{1}{2}\left(\nabla_{\mu}B-m A_{\mu}\right)\left(\nabla^{\mu}B-m A^{\mu}\right) \nonumber \\
&&-\frac{1}{2} \left(m B+\nabla_{\mu} A^{\mu}\right)^2\bigg]+S_{\rm M},
\label{action}
\end{eqnarray}
where $\omega$ is a dimensionless coupling constant, $R$ is the scalar curvature of the spacetime metric $g$, $F^{\mu\nu}$ is the electromagnetic field strength tensor, $A_{\mu}$ and $B$ are vector and scalar fields, respectively. Here the constant $m$ is the mass of the Stueckelberg fields and is defined as a real valued positive number, so that we also avoid an imaginary (tachyonic) mass for the vector field that leads to a ghost instability \cite{Himmetoglu09a,Himmetoglu09b,esposito10}. The action $S_{\rm M}$ stands for the matter source. We use natural units with $\hbar=c=1$ and hence the reduced Planck mass is given by $M_{\rm pl}=1/\sqrt{8\pi G}$, where $G$ is the gravitational coupling. We note that the Stueckelberg action, given in \eqref{action}, preserves gauge invariance under
\begin{equation}
A_{\mu} \rightarrow A_{\mu}+\nabla_{\mu}\lambda\quad \textnormal{and}\quad B \rightarrow B+m\lambda,
\end{equation}
transformations provided that $\lambda$ satisfies
\begin{equation}
(\Box+m^2)\lambda=0.
\end{equation}
Neglecting gravity and investigating in Minkowski spacetime, the action under consideration reduces to the free Stueckelberg action. For free Stueckelberg theory, i.e., for Stueckelberg photon interacting with fermions, the Stueckelberg scalar field $B$ satisfies the free wave equation so that the gauge function $\lambda$ which also satisfies the free wave equation can be used to choose a gauge where $B$ is zero. This is the Proca limit of the Stueckelberg mechanism. However, for our action, in curved spacetime $B$ field does not satisfy the free wave equation so it cannot be set to zero with a gauge transformation.

Now denoting
\begin{eqnarray}
\label{eqn:f}
f=mB+\nabla_{\mu} A^{\mu},
\end{eqnarray}
simplifies the action and varying the action we have
\begin{eqnarray}
\delta S =\int {\rm d}^4x\bigg[&&\delta(\sqrt{-g})\left(-\frac{Rf^2}{8\omega m^2}-\frac{1}{4}F_{\mu\nu}F^{\mu\nu}
+\frac{1}{2}\left(\nabla_{\mu}B-m A_{\mu}\right)\left(\nabla^{\mu}B-m A^{\mu}\right)-\frac{f^2}{2}\right)\nonumber \\
 &&+\sqrt{-g}\bigg(-\frac{R}{4\omega m^2}f\delta f-\frac{f^2}{8\omega m^2}g^{\mu\nu}\delta R_{\mu\nu}-\frac{f^2}{8\omega m^2}\delta g^{\mu\nu}R_{\mu\nu}-\frac{1}{4}\delta\left(F_{\mu\nu} F^{\mu\nu}\right)\nonumber \\
&&+\frac{1}{2}\delta\left(\left(\nabla_{\mu}B-m A_{\mu}\right)\left(\nabla^{\mu}B-m A^{\mu}\right)\right)-f\delta f \bigg)
\;\bigg]+\delta S_{\rm M},
\end{eqnarray}
supplemented by
\begin{eqnarray}\delta{f}=\delta g^{\mu\nu}\nabla_{\nu}A_{\mu}+\nabla_{\mu}(\delta g^{\mu\nu})A_{\nu}-\frac{1}{2}(\nabla^{\alpha}\delta g^{\mu\nu})A_{\alpha}g_{\mu\nu}.\end{eqnarray}
The variations of \eqref{action} with respect to the inverse metric give the Einstein field equations
\begin{eqnarray}
 \frac{f^2}{4\omega m^2}G_{\mu\nu}-\frac{1}{2}g_{\mu\nu}f^2+\frac{1}{4\omega m^2}(g_{\mu\nu}\Box-\nabla_{\mu}
\nabla_{\nu})f^2+\bigg(\frac{R}{4\omega m^2}+1\bigg)(-2\nabla_{\mu}fA_{\nu}+\nabla^{\alpha}A_{\alpha}g_{\mu\nu}f+g_{\mu\nu}\nabla^{\alpha}fA_{\alpha})\nonumber \\
\ar ~~ -\frac{f}{2\omega m^2}A_{\nu}\nabla_{\mu}R+\frac{1}{4\omega m^2}fg_{\mu\nu}A_{\alpha}\nabla^{\alpha}R+
F_\mu ^a F_{\nu a}-\frac{1}{4}g_{\mu\nu}F^{\alpha\beta}F_{\alpha\beta}-(\partial_{\mu}B-mA_{\mu})(\partial_{\nu}B-mA_{\nu}) \nonumber \\ \ar ~~ +\frac{1}{2}g_{\mu\nu}(\partial_{\alpha}B-mA_{\alpha})^2=T_{\mu\nu},
\end{eqnarray}
where $T_{\mu\nu}$ is the energy-momentum tensor of the matter source. The variations of \eqref{action} with respect to the vector field $A_{\mu}$ yield the vector field equation
\begin{eqnarray}
-\frac{1}{4\omega m^2}f\nabla^{\mu}R-\frac{R}{4\omega m^2}\nabla^{\mu}f-\nabla_{\alpha}F^{\alpha\mu}-\nabla^{\mu}(\nabla_{\alpha}A^{\alpha})-m^2A^{\mu}=0.
\end{eqnarray}
Finally we obtain the scalar field equation from the variations of \eqref{action} with respect to the scalar field $B$,
\begin{eqnarray}
\left(\Box+\frac{R}{4\omega}+m^2\right)B+\frac{R}{4\omega m}\nabla_{\mu}A^{\mu}=0.
\end{eqnarray}

We note here that gravitational gauge invariance under general coordinate transformations is also preserved. We consider the spatially flat Robertson-Walker (RW) metric with a maximally symmetric spatial section
\begin{eqnarray}
{\rm d}s^2={\rm d}t^2-a(t)^2[{\rm d}x^2+{\rm d}y^2+{\rm d}z^2],
\end{eqnarray}
where $a(t)$ is the scale factor and $t$ is the cosmic time. Hence, the non-zero components of
the Ricci tensor and the Ricci scalar are given by $R_{00}=-3\ddot{a}/a$, $R_{\alpha \beta }=\left( a%
\ddot{a}+2\dot{a}^{2}\right) \delta _{\alpha \beta }$, $\alpha,\beta =1,2,3$ and $R=-6\left( \ddot{%
a}/a+\dot{a}^{2}/a^{2}\right) $, respectively.

Consistently with the spatially isotropic and homogeneous RW metric, we represent the energy-momentum tensor of the matter source with
\begin{equation}
T_{\nu }^{\mu}={\rm diag}\left[\rho,-p,-p,-p\right],
\end{equation}
where $\rho$ and $p$ are the energy density and pressure respectively and may be functions of cosmic time $t$ only, then we consider spatially homogeneous scalar field
\begin{eqnarray}
B=B(t)
\end{eqnarray}
and finally consider only the scalar potential of the vector field, i.e., spatial part of the vector field is null, as follows:
\begin{eqnarray}
A_0=A(t) \quad\textnormal{and}\quad A_\alpha=0.
\end{eqnarray}

We thus end up with a system of ordinary differential equations given below to be solved:
\begin{eqnarray}
\frac{3f^2}{4\omega m^2}\left(\frac{\dot{a}^{2}}{a^{2}}\right)-\frac{f^2}{2}+\left(\frac{R}{4\omega m^2}+1\right)\left(-\dot fA+f\dot A+3fA \frac{\dot{a}}{a}\right)+\frac{3f \dot f}{2\omega m^2}\frac{\dot{a}}{a}-\frac{fA\dot{R}}{4\omega m^2}-\frac{1}{2}(\dot B-mA)^2=\rho,
\label{rho}
\end{eqnarray}
\begin{eqnarray}
&&-\frac{f^2}{4\omega m^2}\left(2\frac{\ddot{%
a}}{a}+\frac{\dot{a}^{2}}{a^{2}}\right)+\frac{f^2}{2}-\frac{1}{4\omega m^2}\left(2\dot f^2+2f\ddot f+4f\dot{f}\frac{\dot{a}}{a}\right)-\left(\frac{R}{4\omega m^2}+1\right)\left(\dot fA+f\dot A+3fA\frac{\dot{a}}{a}\right) \nonumber \\
&&-\frac{fA\dot{R}}{4\omega m^2}-\frac{1}{2}(\dot B-mA)^2=p,
\label{pressure}
\end{eqnarray}
\begin{eqnarray}
\ddot A+3\dot A\frac{\dot{a}}{a} +3A\left(\frac{\ddot{%
a}}{a}-\frac{\dot{a}^{2}}{a^{2}}\right)+m^2 A+\frac{\dot R f+\dot{f}R}{4\omega m^2}=0,
\label{vector}
\end{eqnarray}
\begin{eqnarray}
\ddot B+3 \dot B \frac{\dot{a}}{a}+m^2 B+\frac{Rf}{4 \omega m}=0,
\label{scalar}
\end{eqnarray}
where
\begin{eqnarray}
\label{notaconstant}
f=mB+\dot{A}+3 A \frac{\dot{a}}{a}.
\end{eqnarray}

We would like to note at this point that the massive Jordan-Brans-Dicke limit cannot be achieved from the action we consider relying on the Stueckelberg theory. At first sight it seems that at $A\rightarrow 0$ limit, in the action only the massive scalar field remains and the $\omega$ becomes the JBD coupling parameter. However, it is well known that substituting $A=0$ in the action is not the same with substituting $A=0$ in the equations of motion. Indeed, one may check that equation \eqref{vector} brings an additional constraint on the system as
 \begin{equation}
\frac{\dot{R}}{R}+\frac{\dot{B}}{B}=0
\end{equation}
for $A\rightarrow 0$ case, hence the solutions that would be obtained with $A\rightarrow 0$ will be different than the massive JBD solutions.

This system is consist of four linearly independent ordinary differential equations \eqref{rho}-\eqref{scalar} that should be satisfied by five unknown functions $\rho$, $p$, $A, B, a$ and therefore is not fully determined. The customary way of determining the system fully at this stage is to introduce an equation of state (EoS) that characterizes the internal properties of the matter source
\begin{equation}
p=w\rho,
\end{equation}
where $w$ is the EoS parameter of the matter source, which is not necessarily constant, but is a constant for the most commonly considered sources in cosmology; namely, takes values $0$, $\frac{1}{3}$ and $-1$ for dust, radiation and cosmological constant respectively. However, the system is far too complicated to be solved analytically and its general solution cannot be obtained even under the assumption of a matter source with a constant EoS parameter. On the other hand, in what follows we shall give various solutions following a strategy moving on from the relation between $f$ and the effective gravitational coupling $G$ that gives us opportunity to investigate some properties of the model that might be of interest from the cosmological point of view.

\section{The Cosmological Solutions}
In comparison with Einstein-Hilbert action of general relativity, the term $f$ in front of the scalar curvature $R$ can be related to the gravitational coupling as follows:
\begin{equation}
\frac{f^2}{8\omega m^2}=\frac{1}{16 \pi G}.
\label{coup}
\end{equation}
We note that, however, in our model $f$ can be time dependent hence it can give rise to a time dependent effective gravitational coupling. Therefore the investigation of our model may be done by considering this property of our model and the constrains on the possible time variation of the effective gravitational coupling utilizing the following relation:
\begin{equation}
\frac{\dot{f}}{f}=-\frac{1}{2}\frac{\dot{G}}{G}
\end{equation}
that follows \eqref{coup}. The constraints on the the rate of change of the gravitational coupling $|\dot{G}/G|$ from various observations (big bang nucleosynthesis, pulsar timing and etc.) can be given as $10^{-10}-10^{-12}\,{\rm yr}^{-1}$. For instance, in a recent study \cite{pulsating} it is given as $\approx -1.8 \times 10^{-10}\,{\rm yr}^{-1}$ from pulsating white dwarfs. One may see \cite{Uzan11} for a comprehensive and recent review on the possible time variation of the effective gravitational coupling. We restrict our study in this paper with the cosmological solutions for which the function $f$ and hence the effective gravitational coupling $G$ are time independent, although it may be possible to obtain solutions with time varying $f$ (hence $G$) consistent with these constraints. However, we do not ignore the possibility of varying effective gravitational coupling and give two sets of solutions: We shall first give solutions for which $f$ is a non-zero constant in subsection \ref{sec:fneq0}. We then give solutions for which $f$ is zero, which corresponds to infinitely large $G$, in subsection \ref{sec:feq0}. We discuss that this extreme case may be considered in the context of very early universe by giving a solution that is compatible with inflationary cosmology.

\subsection{Case I: $f={\rm constant}\neq 0$}
\label{sec:fneq0}
In this case, we assume that the effective gravitational coupling $G$ is a finite positive constant as in general relativity, and hence $w>0$ from \eqref{coup} and $f$ is a finite valued non-zero constant as
\begin{equation}
\label{eqn:fconst}
f=mB+\dot{A}+3 A \frac{\dot{a}}{a}={\rm constant}\neq 0.
\end{equation}
According to this assumption $B$, $A$ and $a$ can still be dynamical but such that $f$ will be yielding a constant value, and the system \eqref{rho}-\eqref{scalar} to be solved reduces to
\begin{eqnarray}
\frac{3f^2}{4\omega m^2}\left(\frac{\dot{a}^{2}}{a^{2}}\right)-\frac{f^2}{2}+\left(\frac{R}{4\omega m^2}+1\right)\left(f\dot A+3fA \frac{\dot{a}}{a}\right)-\frac{fA\dot{R}}{4\omega m^2}-\frac{1}{2}(\dot B-mA)^2=\rho,
\label{rho2}
\\
-\frac{f^2}{4\omega m^2}\left(2\frac{\ddot{%
a}}{a}+\frac{\dot{a}^{2}}{a^{2}}\right)+\frac{f^2}{2}-\left(\frac{R}{4\omega m^2}+1\right)\left( f\dot{A}+3fA\frac{\dot{a}}{a}\right)-\frac{fA\dot{R}}{4\omega m^2}-\frac{1}{2}(\dot B-mA)^2=p,
\label{pressure2}
\\
\ddot A+3\dot A\frac{\dot{a}}{a} +3A\left(\frac{\ddot{%
a}}{a}-\frac{\dot{a}^{2}}{a^{2}}\right)+m^2A+\frac{\dot R f}{4\omega m^2}=0,
\label{vector2}
\\
\ddot B+3 \dot B \frac{\dot{a}}{a}+m^2 B+\frac{Rf}{4 \omega m}=0,
\label{scalar2}
 \end{eqnarray}
supplemented by \eqref{eqn:fconst}. We obtain two different solutions of the system that could be of interest in cosmology.

\subsubsection{Solution I}
\label{sec:sol:1}
In this solution the universe exhibits a de Sitter expansion; the scale factor $a$, Hubble parameter $H$ and the deceleration parameter $q$ of the universe are given as follows:
\begin{eqnarray}
a=a_{1}e^{\sqrt{\frac{\omega}{3}}mt},\quad H=\frac{\dot{a}}{a}=\sqrt{\frac{\omega}{3}}m\quad\textnormal{and}\quad q=-\frac{\ddot{a}a}{\dot{a}^2}=-1,
\end{eqnarray}
where $a_{1}$ is the integration constant. We find that the scalar field is a constant and the vector field is null
\begin{eqnarray}
\label{eqn:fieldssol1}
B=\frac{f}{m}\quad \textnormal{and}\quad A=0.
\end{eqnarray}
The energy density and pressure of the matter source are found to be constant as follows
\begin{eqnarray}
\label{matlambda}
p=-\rho=\frac{f^2}{4}.
\end{eqnarray}

The universe expands exponentially with a rate directly proportional to $m$, and is static for $m=0$. This is a result in line with our expectation that there may be a connection between the accelerated expansion of the universe and the small but non-zero mass term of the Stueckelberg fields. The matter source predicted in this solution \eqref{matlambda}, on the other hand, yields an EoS in the form of a cosmological constant and a negative energy density with a particular value. A source with a negative energy density that does not violate the dominant energy condition, which implies that energy does not flow faster than the speed of light, is allowed only if it is in the form of vacuum energy. Accordingly, adding a bare cosmological constant $\bar{\Lambda}$ to the action \eqref{action} as
\begin{equation}
\label{shiftaction}
S\rightarrow S-\bar{\Lambda}\int \sqrt{-g}\; {\rm d}^{4}x,
\end{equation}
the energy density and pressure of the matter source given in \eqref{rho} and \eqref{pressure}, and hence given in \eqref{rho2} and \eqref{pressure2}, will be shifted as
\begin{equation}
\label{shift}
\rho\rightarrow \rho+\bar{\Lambda} \quad\textnormal{and}\quad p\rightarrow p-\bar{\Lambda},
\end{equation}
while the equations of the vector and scalar fields  \eqref{vector} and \eqref{scalar}, and hence \eqref{vector2} and \eqref{scalar2}, are unchanged. Therefore, the energy density and pressure of the matter source given in \eqref{matlambda} can now be elevated to zero,
\begin{equation}
p=0=\rho,
\end{equation}
by choosing
\begin{equation}
\label{lambdashft}
\bar{\Lambda}=-\frac{f^2}{4},
\end{equation}
which is always negative since $f$ is a non-zero real number. We note that $\bar{\Lambda}$ indeed corresponds to the energy density of the vacuum, i.e. $\bar{\Lambda}=\rho_{\rm vac}$ and in this sense it is not the cosmological constant defined by $\Lambda=8\pi G \rho_{\rm vac}$. The reason being that the effective gravitational coupling in the action we considered at the beginning \eqref{action} is not necessarily constant as can be seen from \eqref{eqn:f}, and hence, in contrast to GR, adding a cosmological constant as $R\rightarrow R-2\Lambda$ to the action would not correspond to adding a vacuum energy. Negative vacuum energies, on the other hand, appear in string theory (and other models of quantum gravity), supersymmetry, super gravity and etc. and have been largely studied for addressing the cosmological constant problem \cite{Sahni00,Nobbenhuis06}. For instance, in exact supergravity the lowest energy state of the theory, generically has negative energy density \cite{Nobbenhuis06} and string theory, the most prominent candidate for a consistent theory of quantum gravity, naturally predicts the existence of negative energy vacua \cite{Kachru03}. This introduction of a negative vacuum energy with a particular energy density for elevating the energy density of the matter source to zero will particularly be very useful in the investigation of the following solution.

\subsubsection{Solution II}

In this solution the universe starts expanding with a decelerated expansion rate and then starts to accelerate at a certain time; setting $a=0$ at $t=0$, we obtain the scale factor, Hubble parameter and deceleration parameter as follows:
\begin{eqnarray}
\label{eqn:fcsol2}
a&=&a_1\sinh^{1/2}{\left(2\sqrt{\frac{w}{3}}mt\right)},\quad H=\sqrt{\frac{\omega}{3}} m\coth{\left(2\sqrt{\frac{w}{3}}mt\right)}\quad\textnormal{and}\quad q=8\cosh^2{\left(2\sqrt{\frac{w}{3}}mt\right)}-1,
\end{eqnarray}
where $a_{1}$ is the integration constant. We find that the scalar field is a constant and the vector field is null
\begin{equation}
B=\frac{f}{m} \quad\textnormal{and}\quad A=0.
\end{equation}
The energy density and pressure of the matter source are obtained as follows:
\begin{equation}
\label{rhobizz}
\rho=-\frac{f^2}{4}+f^2  \sinh^{-2}{\left(2\sqrt{\frac{w}{3}}mt\right)}\quad\textnormal{and}\quad p=\frac{f^2}{4}+\frac{f^2}{3}\sinh^{-2}{\left(2\sqrt{\frac{w}{3}}mt\right)},
\end{equation}
that yield the following EoS parameter
\begin{equation}
w=\frac{3+4\sinh^{-2}{\left(2\sqrt{\frac{w}{3}}mt\right)}}{-3+12  \sinh^{-2}{\left(2\sqrt{\frac{w}{3}}mt\right)}}.
\end{equation}

We note first that the scale factor has a similar behavior with the $\Lambda$CDM model with the difference that they have different powers; it is $\frac{1}{2}$ in this solution while it is $\frac{2}{3}$ in the $\Lambda$CDM model. In the $\Lambda$CDM model, which is based on GR, the universe evolves from pressure-less matter ($w=0$) dominated universe to $\Lambda$ dominated universe (de Sitter universe), such that $q\sim \frac{1}{2}$ at $t\sim 0$ and $q\rightarrow -1$ as $t\rightarrow \infty$. One may check that, on the other hand, solving field equations in GR in the presence of $\Lambda$ and radiation/relativistic fluid, which can be described with an EoS parameter $w=1/3$, instead of pressure-less matter, one would obtain the same behavior we obtained for the scale factor \eqref{eqn:fcsol2} in this solution, which yields $q\sim 1$ at $t\sim 0$ and $q\rightarrow -1$ as $t\rightarrow \infty$. In GR, the value $q=1$ corresponds to the value of the deceleration parameter in the radiation dominated universe that can describe the early universe, e.g., the time when primordial nucleosynthesis took place. We note that the fluid we obtained in this solution also has the EoS parameter equal $\frac{1}{3}$ at $t=0$ but  exhibits a bizarre behavior later on; it reaches infinitely large positive values at $t_{\rm c}=\frac{1}{m}\sqrt{\frac{3}{2w}}\ln(3+2\sqrt{2})$, and then starts with an infinitely large negative value at  $t_{\rm c}$ and approaches monotonically to $-1$ as $t\rightarrow\infty$. The reason being that its energy density becomes zero and changes sign at $t_{\rm c}$ and then approaches a negative constant equal to $-\frac{f^2}{4}$ as $t\rightarrow\infty$, all the while the pressure decreases too but at a slower rate and approaches a positive constant equal to $\frac{f^2}{4}$ as $t\rightarrow\infty$. In fact, one may check that, as $t\rightarrow\infty$, this solution approaches the solution we gave above in section \ref{sec:sol:1}, where we elevate the energy density of the matter source to zero by introducing a negative vacuum energy density with a value equal to $-\frac{f^2}{4}$. Let us now apply the same procedure \eqref{shiftaction}; using equations \eqref{shift} and \eqref{lambdashft}, namely introduce a vacuum energy with an energy density equal to $-\frac{f^2}{4}$, the energy density and pressure of the matter source given in \eqref{rhobizz}  can now be written as follows:
\begin{equation}
\rho=f^2  \sinh^{-2}{\left(2\sqrt{\frac{w}{3}}mt\right)} \quad\textnormal{and}\quad  p=\frac{f^2}{3}\sinh^{-2}{\left(2\sqrt{\frac{w}{3}}mt\right)}
\end{equation}
yielding the following properties
\begin{equation}
\rho\propto a^{-4} \quad\textnormal{and}\quad w=\frac{1}{3},
\end{equation}
which is exactly the EoS that describes radiation/relativistic fluid. It is interesting that this solution obtained by assuming that $f$ is constant, hence effective gravitational coupling is constant, doesn't predict an unknown kind of matter source but a radiation/relativistic fluid provided that the negative vacuum energy density is isolated appropriately.

\subsection{Case II: $f=0$}
\label{sec:feq0}
As we mentioned previously, in this paper we restrict the investigation of the model with the cases for which $f$ is constant that gives rise to a time independent effective gravitational constant. Now, in this section, we investigate an extreme case for a constant $f$ solution such that
\begin{equation}
\label{eqn:f0}
f=mB+\dot{A}+3 A \frac{\dot{a}}{a}=0,
\end{equation}
which corresponds to an infinitely large effective gravitational coupling limit. Although such an extreme case may not be advocated as a physically viable case, an investigation of the solution under this assumption may give us an idea about the behavior of our model in case of very large values of the effective gravitational coupling. Although there are strong constraints on the possible time variation of the gravitational coupling in the observable past of the universe, our understanding on the very early universe, strictly speaking the time scales between the Planck time scale $10^{-43}$ s and SUSY breaking time scale $<10^{-10}$ s, is still quite speculative. Indeed there is no fundamental theory of physics that assures the constancy of the gravitational coupling at the energy scales that correspond to the time scales close to the Planck time scales. Hence, there is a room for the solutions that are obtained in this extreme case, such that they maybe considered in the context of the dynamics of the very early universe, for instance, in the context of inflation that is believed to took place at time scales $\sim 10^{-35}$ s with the corresponding energy scales $\sim10^{15}$ GeV. It is also noteworthy to point out here that setting $f$ equal to zero in our theory described by the action given in \eqref{action} is in fact not the same as setting a constant of a theory to zero, namely, as setting inverse of the gravitational coupling constant $1/G$ to zero in GR described by Einstein-Hilbert action: $f$ is in fact not a true constant of our model/the action \eqref{action} but a dynamical parameter consisting of three additive terms that are dynamical too (see \eqref{notaconstant}). Hence, the investigation of a solution under the assumption $f=0$ should be understood as the investigation of the behavior of our model in the period of time when the constituents of $f$ possibly evolve such that $f$ vanishes.

In this case, i.e., choosing \eqref{eqn:f0}, equations \eqref{rho}-\eqref{scalar} reduce to the following
\begin{eqnarray}
\label{rhof0}
-\frac{1}{2}(\dot B-m A)^2=\rho,\\
\label{pf0}
-\frac{1}{2}(\dot B-m A)^2=p,\\
\label{vecf0}
\ddot A+3\dot{A} \frac{\dot{a}}{a} +3A \left(\frac{\ddot{a}}{a}-\frac{\dot{a}^2}{a^2}\right)+m^2 A=0,\\
\label{sclr0}
\ddot B+3\dot{B} \frac{\dot{a}}{a}+m^2 B=0,
\end{eqnarray}
supplemented by \eqref{eqn:f0}. It is important to note here that this reduced system \eqref{eqn:f0}-\eqref{sclr0} is not fully determined since there are five unknown functions but only four linearly independent equations in this case: Differentiating \eqref{eqn:f0} once and using the result in \eqref{vecf0} we find
\begin{equation}
\label{eqn:mb}
mA=\dot{B},
\end{equation}
and substituting this back into \eqref{eqn:f0} we get \eqref{sclr0}, which means that we lost one equation and hence one additional constraint is required to fully determine the system. We first give the solution of this undetermined system in terms of the ratio of the vector and scalar fields denoted as
\begin{equation}
\label{eqn:F}
F=\frac{A}{B},
\end{equation}
which will provide us with an insight for choosing a useful and reasonable function for the additional constraint rather than an arbitrary function. Now using \eqref{eqn:f0} we obtain the scale factor as
\begin{equation}
\label{eqn:inf1}
a=a_1 e^{-\frac{1}{3}\int m\frac{B}{A}+\frac{\dot{A}}{A}\;{\rm d}t},
\end{equation}
where $a_1$ is an integration constant. Next using \eqref{eqn:F} with \eqref{eqn:mb} and \eqref{eqn:inf1} we find that the scale factor $a$, the scalar field $B$ and the vector field $A$ can be written in terms of $F$ as
\begin{equation}
\label{eqn:f0gen}
a=a_1 e^{-\frac{1}{3} \int \frac{m}{F}+\frac{\dot{F}}{F}+mF\; {\rm d}t}, \quad B=B_{1} e^{\int mF\; {\rm d}t} \quad\textnormal{and}\quad  A=F B_{1} e^{\int mF\; {\rm d}t}
\end{equation}
where $B_{1}$ is an integration constant. The energy density and pressure, on the other hand, are always null as can immediately be seen upon substituting \eqref{eqn:mb} in \eqref{rhof0} and \eqref{pf0}
\begin{equation}
\label{eqn:inf2}
\rho=0\quad\textnormal{and}\quad p=0,
\end{equation}
i.e., there is nothing in the universe other than the vector and scalar fields, which is plausible since the presence of a matter source in this extreme case would be fatal.

We note that the scale factor given in \eqref{eqn:f0gen} possesses some interesting properties. To make this more clear, we give also the Hubble and deceleration parameters in terms of $F$:
\begin{equation}
\label{eqn:f0Hgen}
3H=-\frac{m}{F}-\frac{\dot{F}}{F}-mF\quad\textnormal{and}\quad q=3\frac{-m\dot{F}+F\ddot{F}-\dot{F}^2+m\dot{F}F^2}{(m+\dot{F}+mF^2)^2}-1.
\end{equation}
We note that the Hubble parameter consists of three additive terms: Two terms that contribute to the Hubble parameter positively if the vector and scalar fields yield opposite signs ($F<0$). These are directly proportional to the mass term $m$, and while one of them is directly proportional to $F$, the other is inversely proportional to $F$. And another term ($-\dot{F}/F$) that contributes to the Hubble parameter positively/negatively if the rate of change of the vector field is less/higher than that of the scalar field. In contrast to the other two, this term is independent of the mass term and arises only if the ratio between the scalar and vector fields is not constant. Accordingly, the expansion of the universe, viz. the Hubble parameter, is not only contributed by the distinct behaviors of the vector and scalar fields ($-\dot{F}/F$), but also, interestingly, by the ratio of these two fields in a non-trivial way ($-m(1/F+F)$) if there is a non-zero mass term. It is apparent that the presence of a non-zero mass term can lead to an intricate expansion history of the universe, even if the evolution of the ratio between the scalar and vector fields obeys a simple function. On the other hand, if $m$ is null and/or $F$ is constant then we obtain the following simple cases:
\begin{itemize}
\item
If the mass term is non-zero and the ratio between the scalar and vector fields is constant then we have:
\begin{equation}
H=-\frac{m}{3F}-\frac{mF}{3} \quad\textnormal{and}\quad   A\propto B\propto \exp(mFt)  \quad (m>0\textnormal{ and } F={\rm const.}).
\end{equation}
If $F<0$, the universe exhibits de Sitter expansion and the scalar and vector fields decrease exponentially as $t$ increases.
\item
If the mass term is null and the ratio between the scalar and vector fields is not constant then we have
\begin{equation}
a\propto A^{-\frac{1}{3}} \quad\textnormal{and}\quad  B={\rm const.} \quad \quad (m=0\textnormal{ and } F\neq{\rm const.}).
\end{equation}
\item
If the mass term is null and the ratio between the scalar and vector fields is a constant then we have a static universe:
\begin{equation}
a={\rm const.},\quad B={\rm const.} \quad\textnormal{and}\quad A=0  \quad (m=0\textnormal{ and } F={\rm const.}).
\end{equation}
\end{itemize}
In the light of the above discussion, let us now determine the equations given in \eqref{eqn:f0gen} by making a plausible assumption on the time evolution of the ratio between the scalar and vector fields, i.e., $F$. We demand (i) the universe to start from a singularity at $t=0$, namely, $H\rightarrow+\infty$ and $a\rightarrow0$ as $t\rightarrow0$, which can be achieved if either $F\rightarrow0$ as $t\rightarrow0$ or $F\rightarrow-\infty$ as $t\rightarrow0$, as can be seen from \eqref{eqn:f0Hgen}, (ii) the assumed function for $F$ to yield minimum number of free parameters, namely only one, but yet can realize the simple cases given above as particular cases as well as various cases depending on the value of the free parameter, (iii) the model to be able to approximate a power-law expansion (i.e., $H\propto t^{-1}$) for a certain period of time, which maybe achieved due to the term $\frac{\dot{F}}{F}$ in \eqref{eqn:f0Hgen}. The simplest function that can be utilized in accordance with all our demands is maybe a power-law relation given as follows
\begin{equation}
\label{eqn:asmpf0}
F=\frac{A_{0}}{B_{0}} \left(\frac{t}{t_{0}}\right)^{-k},
\end{equation}
where $k$ is a constant whose sign will determine whether the vector field will be dominant over the scalar field at the earlier times or the later times. Finally, solving \eqref{eqn:inf1} using this assumption \eqref{eqn:asmpf0} we obtain the scale factor as
\begin{subequations}
\label{eqn:infsolf}
\begin{eqnarray}
\label{eqn:infsol}
a=a_{0}t^{\frac{k}{3}}
\times
\exp\left[\frac{A_{0}}{B_{0}}\frac{m t_{0}}{3(k-1)} \left(\frac{t}{t_{0}}\right)^{-k+1}\right]
\times
\exp\left[-\frac{B_{0}}{A_{0}}\frac{m t_{0}}{3(k+1)}\left(\frac{t}{t_{0}}\right)^{k+1}\right] \quad &\textnormal{for}& \quad |k| \neq 1,
\\
a = a_0 t^{-\frac{B_{0}}{A_{0}}\frac{mt_{0}}{3}-\frac{1}{3}} \times \exp\left[-\frac{A_{0}}{B_{0}}\frac{mt_{0}}{6}\left(\frac{t}{t_{0}}\right)^2\right]\quad &\textnormal{for}& \quad k=-1,
\\
a = a_0 t^{-\frac{A_{0}}{B_{0}}\frac{mt_{0}}{3}+\frac{1}{3}} \times \exp\left[-\frac{B_{0}}{A_{0}}\frac{mt_{0}}{6}\left(\frac{t}{t_{0}}\right)^2\right]\quad &\textnormal{for}& \quad k=1.
\end{eqnarray}
\end{subequations}
On the other hand, one may check that the Hubble and deceleration parameters can be given uniquely for arbitrary values of $k$ as
\begin{equation}
\label{eqn:hinf}
H=-\frac{A_{0}}{B_{0}} \frac{m}{3} \left(\frac{t}{t_{0}}\right)^{-k}    +\frac{k}{3}t^{-1}     -\frac{B_{0}}{A_{0}}\frac{m}{3}  \left(\frac{t}{t_{0}}\right)^{k},
\end{equation}
\begin{equation}
q=3k{t_{0}}^{k}t^{k-1}\frac{A_{0}}{B_{0}} \frac{\frac{A_{0}}{B_{0}} {t_{0}}^{k}  t^{k-1} + m(t^{2k}-\frac{{A_{0}}^{2}}{{B_{0}}^{2}} {t_{0}}^{2k})}{\left[  k \frac{A_{0}}{B_{0}} {t_{0}}^{k}  t^{k-1} - m(t^{2k}+\frac{{A_{0}}^{2}}{{B_{0}}^{2}} {t_{0}}^{2k})   \right]^2}-1.
\end{equation}
We will show that the three terms in the expression \eqref{eqn:hinf} dominate in different eras giving an inflationary phase followed by a deceleration phase followed by an acceleration phase provided that the values of the constants are chosen appropriately.
We obtain the scalar and vector fields as follows:
\begin{subequations}
\begin{eqnarray}
B=B_{0} e^{-\frac{A_{0}}{B_{0}}\frac{mt_{0}}{k-1}\left(\frac{t}{t_{0}}\right)^{-k+1}}
\quad\textnormal{and}\quad
A=A_{0} \left( \frac{t}{t_{0}}    \right)^{-k} e^{-\frac{A_{0}}{B_{0}}\frac{mt_{0}}{k-1}\left(\frac{t}{t_{0}}\right)^{-k+1}}\quad &\textnormal{for}& \quad k \neq 1,
\\
B=B_{0} \left(\frac{t}{t_{0}}\right)^{\frac{A_{0}}{B_{0}}mt_{0}}
\quad\textnormal{and}\quad
A=A_{0} \left(\frac{t}{t_{0}}\right)^{\frac{A_{0}}{B_{0}}mt_{0}-1}\quad &\textnormal{for}&\quad k=1.
\end{eqnarray}
\end{subequations}
We note that the evolution of the scale factor \eqref{eqn:infsolf} is characterized by the mass term $m>0$ (in particular, according to whether it is null or non-null) and the constant $k$ that determines the relative rate of change of the scalar and vector fields with respect to time \eqref{eqn:asmpf0}. Hence, in what follows, we shall carry out a detailed discussion considering the cases $m=0$ and $m\neq 0$ separately.

\subsubsection{The case $m=0$: Power-law expansion}

We note that the choice $m=0$ sets the exponential terms to unity and leads to a simple power-law expansion/contraction as
\begin{equation}
a=a_{0}t^{\frac{k}{3}},\quad H=\frac{k}{3}t^{-1}\quad\textnormal{and}\quad q=\frac{3}{k}-1
\end{equation}
with a constant scalar field but a vector field yielding a power-law evolution in time
\begin{equation}
B=B_{0}\quad\textnormal{and}\quad A=A_{0} \left( \frac{t}{t_{0}} \right)^{-k}.
\end{equation}
The vector field is inversely proportional to the volume of the universe $A\propto a^{-3}$, and the universe expands at an accelerating rate if $k>3$ and at a decelerating rate if $0<k<3$ while the universe contracts if $k<0$. The case $k=0$ is a special case for which the universe becomes static and both scalar and vector fields are also constant.

\subsubsection{The case $m\neq 0$: Inflation with a switch-off mechanism}

We showed, in the previous subsection, that the case with zero mass $m=0$ leads to a simple power-law behavior of the scale factor and that the further choice $k=0$ leads to a static universe. We note that the static universe arises since, in equation \eqref{eqn:infsol}, the choice $m=0$ sets the exponential terms to unity while the choice $k=0$ sets the power term to unity. Hence, in this subsection, we shall first consider the case $m\neq 0$ but $k=0$ and then discuss the case $m\neq 0$ and $k\neq 0$ that can give rise to an evolution that might be considered in the context of the inflation mechanism.

We observe that, setting
\begin{equation}
k=0,
\end{equation}
the universe exhibits exponential behavior as
\begin{equation}
a=a_{0} e^{-\left(\frac{A_{0}}{B_{0}}+\frac{B_{0}}{A_{0}} \right) \frac{m}{3}t},\quad H=-\frac{m}{3}\left( \frac{A_{0}}{B_{0}}+\frac{B_{0}}{A_{0}}  \right) \quad\textnormal{and}\quad q=-1,
\end{equation}
and that the scalar and vector fields evolve with the same rate as
\begin{equation}
B=B_{0} e^{\frac{A_{0}}{B_{0}}mt}\quad\textnormal{and}\quad A=A_{0} e^{\frac{A_{0}}{B_{0}}mt}.
\end{equation}
The universe expands exponentially for $m>0$ and $A_{0}/B_{0}<0$ and the value of the Hubble parameter is proportional with the mass term, and hence a static universe is obtained when $m=0$ as expected. This is because the choice $k=0$ in \eqref{eqn:infsol} sets the power term to unity and the exponents of the two exponential terms identically to $t$. On the other hand, the values $k\neq 0$ not only give rise to a power term, but also cause the exponents of the two exponential terms to differ from each other and therefore the power term (dependent on $k$ only) and the two exponential terms (which arise when the mass term is non-zero and are dependent on $k$ in distinct ways) all together give rise to a non-trivial evolution that can even be related with the inflation model.

One may check that the model can give rise to various behaviors depending on the choice of the parameters. However, we are particularly interested in whether the model can give rise to a behavior that is compatible with the inflationary cosmology. Looking at the Hubble parameter \eqref{eqn:hinf} and the scale factor \eqref{eqn:infsol}, it can be easily seen that choosing the values of the parameters appropriately under the assumption $A_{0}/B_{0}<0$ and $k>1$ the universe starts expanding at $t=0$ and will always expand passing through three different stages respectively;
\begin{equation}
\label{phaseI}
 a\sim \exp\left[\frac{A_{0}}{B_{0}}\frac{m t_{0}}{3(k-1)} \left(\frac{t}{t_{0}}\right)^{-k+1}\right],
 \quad
 H\sim -\frac{A_{0}}{B_{0}} \frac{m}{3} \left(\frac{t}{t_{0}}\right)^{-k}
 \quad\textnormal{and}\quad
 q\sim -\frac{3k}{mt_{0}}\frac{B_{0}}{A_{0}}\left(\frac{t}{t_{0}}\right)^{(k-1)}-1\quad \textnormal{at} \quad t\simeq 0,
\end{equation}
then
\begin{equation}
\label{phaseII}
 a\sim t^{\frac{k}{3}},
 \quad
 H\sim \frac{k}{3}t^{-1}
 \quad\textnormal{and}\quad
 q\sim \frac{3}{k}-1\quad \textnormal{at}\quad t\gtrsim 0,
\end{equation}
and finally at later times
\begin{equation}
\label{phaseIII}
 a\sim \exp\left[-\frac{B_{0}}{A_{0}}\frac{m t_{0}}{3(k+1)}\left(\frac{t}{t_{0}}\right)^{k+1}\right],
 \quad
 H\sim -\frac{B_{0}}{A_{0}}\frac{m}{3}  \left(\frac{t}{t_{0}}\right)^{k}
 \quad\textnormal{and}\quad
 q\sim \frac{3k}{mt_{0}}\frac{A_{0}}{B_{0}}     \left( \frac{t}{t_{0}}\right)^{-k-1}-1\quad \textnormal{at}\quad t\gg0.
\end{equation}
In the first stage given by \eqref{phaseI}, the universe begins with an accelerating expansion rate, such that $a\rightarrow 0$, $H\rightarrow\infty$ and $q\rightarrow -1$ as $t\rightarrow 0$. After a while  the power term will become dominant over the two exponential terms in \eqref{eqn:infsol} and the second stage in which the evolution of the universe can be described by \eqref{phaseII} will start. Accordingly, one may check from \eqref{phaseII} that if $1<k<3$ then the accelerated expansion achieved in the previous stage will end and the universe will enter into a decelerated expansion phase, otherwise, i.e. if $k>3$, it will keep on accelerated expansion accordingly \eqref{phaseII}. Eventually, the exponential term on the right will be dominant over the exponential term at the middle and the power term in \eqref{eqn:infsol} and the third stage, in which the universe will be described by \eqref{phaseIII}, will start. Accordingly, the universe will first evolve into a super-accelerated phase ($q<-1$) and eventually will start to approach monotonically to an expansion rate with a deceleration parameter equal $-1$, $a\rightarrow\infty$ and $q\rightarrow-1$ as $t\rightarrow\infty$. In this picture, the case $A_{0}/B_{0}< 0$ with $1<k<3$ is of particular interest since it can give rise to a behavior compatible with inflationary cosmology, such that the expansion of the universe starts with an accelerated expansion that will be switched off and the universe will enter into a decelerated expansion phase. Moreover, interestingly, this decelerated expansion phase will be followed by an another accelerated expansion phase that may be related with the late time acceleration of the universe. Such a behavior, two different accelerated expansion phases with a decelerated expansion phase between them is consistent with the current paradigm in cosmology ($\Lambda$CDM cosmology supplemented by inflationary cosmology).

We demonstrate the evolution of the universe in this solution by giving some suitable values to the parameters. To do so, we first choose $k=\frac{3}{2}$ so that in the decelerated expansion phase that follows the first accelerated expansion phase the value of the deceleration parameter will be $q=1$, which is the value of the deceleration parameter when the primordial nucleosynthesis took place ($\sim 10^2$ seconds after the Big Bang) in the standard cosmology based on GR. We choose $t_{0}=14$ Gyr, $H_{0}=10^{-32}$ eV and $\frac{A_{0}}{B_{0}}=-10^{-13}$ for the present universe, and choose $m=10^{-45}$ eV, which is a value almost 20 orders of magnitude less than the most strict upper limits given for the photon mass. Using these values we find that $q\simeq -1$, $H\simeq 10^{39}\;{\rm s}^{-1}$ and $A/B\simeq-10^{70}$ at $t=10^{-38}\;{\rm s}$ (inflation), $q\simeq 0$, $H\simeq 10^{35}\;{\rm s}^{-1}$ and $A/B=-10^{66}$ at $t=10^{-35}\;{\rm s}$ (inflation ends). In a short while following the end of the inflationary phase, the universe achieves an expansion rate with a deceleration parameter equal to unity and preserves this value for a long time: $q\simeq 1$, $H\simeq 10^{31}\;{\rm s}^{-1}$ and $A/B\simeq -10^{61}$ at $t\simeq 10^{-32}\;{\rm s}^{-1}$, $q\cong 1$, $H\simeq 0.5\;{\rm s}^{-1}$ and $A/B\simeq -10^{13}$ at $t\simeq 1\;{\rm s}$, $q\cong 1$, $H\simeq 0.005\;{\rm s}$ and $A/B\simeq -10^{10}$ at $t\simeq 100\;{\rm s}$. The value of the deceleration parameter does not deviate from the value $q\cong 1$ till the age of the universe reaches $t\simeq 10^{17}\;{\rm s}$. Because the value of the effective gravitational coupling is infinitely large in this solution extending the model for large $t$ values may not be reliable. On the other hand, because there is no matter source ($\rho=0$) in this solution, extending the model to large $t$ values will still be consistent within the model itself. Interestingly, we find that $q\sim -0.4$, $H\simeq 10^{-18}\;{\rm s}^{-1}$ and $A/B\sim -10^{-13}$ at $t\sim 10\; {\rm Gyr}$ and the values of the deceleration and Hubble parameters here are consistent with the observations. We note that the universe in our model begins already with an accelerated expansion rate. Therefore we are not able to calculate the e-fold of the size of the universe between the switch-on and -off of the inflation as in the usual inflationary models. However we calculate in our model that the size of the universe ($a$) goes through 50 e-folds from $t=10^{-38}\;{\rm s}$ to the end of inflation at $t=10^{-35}\;{\rm s}$. As the final remark, we note that the vector field is dominant over the scalar field in the early times, namely, $|A/B|>10^{66}$ when the inflation took place $t<10^{-35}\;{\rm s}$ and $|A/B|\sim 10^{10}$ at $t\sim100\;{\rm s}$, while the scalar field is dominant over the vector field at the times of the late time acceleration, namely, $|A/B|\sim10^{-13}$ at $t=10\;{\rm Gyr}$. This tells us that it is the vector field who is responsible for the inflationary phase in the early universe while it is the scalar field who is responsible for the late time acceleration of the universe.

\section{Concluding remarks}

Introducing a mass for a vector field, e.g., a mass to a photon, requires reorganization of the degrees of freedom. The mechanism to achieve this by preserving the gauge symmetry is known as the Stueckelberg mechanism. Proposing an extra scalar field as an extra degree of freedom seems similar to the usage of the JBD field in cosmology, we have introduced an action by extending this idea and constructed a cosmological model in Robertson-Walker spacetime. We have also showed that this model does not reduce to cosmological model in massive JBD theory for zero vector field.

The effective gravitational coupling in the model is determined by three dynamical parameters; the scalar and vector fields as well as the expansion rate of the universe. We have given expanding universe solutions under the assumption that the effective gravitational coupling is constant, which implies that the scalar and vector fields can be dynamical but subject to the invariability of the effective gravitational coupling. We have given two sets of solutions: the case $f$ is constant and nonzero, which is the case similar to GR with constant gravitational coupling, and the case $f=0$, which is an extreme case that corresponds to infinitely large effective gravitational coupling.

In the case $f={\rm constant}\neq0$, the universe is static if the mass term of the Stueckelberg fields is null. If the mass term has a positive real value, then the universe exhibits either a de Sitter expansion or a $\Lambda$CDM type expansion but with a different power. We showed that these two solutions predict a certain amount of negative vacuum energy, and that while the former solution is matter source-free, the latter solution  involves radiation/relativistic fluid. In the case $f=0$, we have found a matter source free solution which can yield a behavior compatible with the inflationary cosmology (including the switch-off mechanism) provided that the mass term is positive valued, unless otherwise it gives nothing but a simple power law expansion. In particular, we obtain a universe going through a deceleration phase sandwiched by two different accelerated expansion phases provided that the vector field decays faster than the scalar field as the universe expands, which in turn implies that essentially the vector field $A$ drives the inflation while the scalar field $B$ gives rise to a late time acceleration. Moreover the solution allows us to set the value of the deceleration parameter to a value required for a successful primordial nucleosynthesis and the decelerating expansion phase can last for long enough time. However, although this solution gives very interesting dynamics for the universe, the effective gravitational coupling yielding infinitely large values stands as an important issue to be faced. We think that solutions giving rise to such an interesting behavior of the universe but not suffering from this issue may be obtained by allowing the effective gravitational coupling ($G=wm^2/2\pi f^2$) to be a particular function of time such that it will start with infinitely large values but will then approach to a non-zero value by changing slowly enough after the end of inflation consistently with the observational constraints. We are working for such solutions as the extension of this work and our results will be reported elsewhere.

 \section{Acknowledgments}
\"{O}.A. acknowledges the support by T\"{U}B{\.I}TAK Research Fellowship for Post-Doctoral Researchers (2218). \"{O}.A. appreciates also the support from Ko\c{c} University. M.A. acknowledges the support of Turkish Academy of Sciences and also the support of Bo\u{g}azi\c{c}i University Scientific Research (BAP) project number $6324$. N.K. thanks Bo\u{g}azi\c{c}i University for the financial support provided by the Scientific Research (BAP) Project number $7128$.
M.K. acknowledges the support of Bo\u{g}azi\c{c}i University Scientific Research (BAP) Project number $6700$.

\end{document}